# Pervasive Communications Technologies For Managing Pandemics

Mohammad Ilyas
College of Engineering and Computer Science
Florida Atlantic University
Boca Ratom, FL 33431 USA
ilyas@fau.edu

Basit Qureshi
College of Computer and Information Sciences
Prince Sultan University
Riyadh, 11586 Saudi Arabia
qureshi@psu.edu.sa

*Abstract*—Pandemics always have had serious consequences unless they were effectively contained. Recent experiences with COVID-19 show that by using a smart and swift approach to deal with pandemics, avoids overwhelming of healthcare systems, and reduces loss of precious life. This paper is about using smart technologies such as Mobile Edge Clouds (MEC), Internet of Things (IoT), and Artificial Intelligence (AI), as an approach to effectively manage pandemics. IoT provides pervasive connectivity among various devices and can be used for collecting information such as location and symptoms of potentially infected individuals. MECs provide cloud services on the edge, integrating IoT infrastructure and execution of sophisticated AI algorithms in the Cloud. In this paper, we develop a prototype to demonstrate the convergence of pervasive technologies to support research in managing pandemics. Low-cost Single Board Computers (SBC) based clusters are integrated within MEC to support remote medical teams in the field. The prototype implements a lightweight Docker container orchestrated by Kubernetes eco-system which is deployed on the clusters. The prototype successfully demonstrates that mobile medical facilities can utilize the proposed solution to collect information and execute AI algorithms while on the go. Finally, we present a discussion on the role of converging pervasive technologies on managing pandemics.

*Keywords—pandemic management, Internet of things (IoT), artificial intelligence (AI), Mobile Edge Clouds, Kubernetes, Single Board Computer Clusters.*

## I. Introduction

History of pandemics is as long as the history of humanity. Although not all details on earlier pandemics are documented, there is some information available on the major pandemics dating back to 5000 years [1].

- Prehistoric epidemic: Circa 3000 B.C.
- Plague of Athens: 430 B.C.
- Antonine Plague: 165 – 180 A.D.
- Plague of Cyprian: 250 – 271 A.D.
- Plague of Justinian: 541 – 542 A.D.
- The Black Death: 1346 – 1353 A.D.
- Great Plague of London: 1665 – 1666 A.D.
- Great Plague of Marseille: 1720 – 1723 A.D.
- Russian plague: 1770 – 1772 A.D.
- Philadelphia yellow fever epidemic: 1793 A.D.
- American polio epidemic: 1916 A.D.
- Spanish Flu: 1918 – 1920 A.D.
- Asian Flu: 1957 – 1958 A.D.
- AIDS pandemic and epidemic: 1981 – Present
- H1N1 Swine Flu pandemic: 2009 – 2010 A.D.
- West African Ebola epidemic: 2014 – 2016 A.D.
- Zika Virus epidemic: 2015 – Present
- COVID-19 Novel Coronavirus: 2019 – Present

The global population is currently over 7.5 Billion and is rapidly increasing. Over 4.1 Billion people live in urban areas, often in densely populated cities and towns. The congested urban areas and dense population centers provide a hotbed for pandemics to flourish and have a devastating effect on population health and the resulting death toll [2]. The recent novel coronavirus COVID-19 pandemic rapidly spread throughout the globe. Despite the availability of advanced diagnostic and treatment options, the global infection rate and the death toll due to COVID-19 is still very high. Thanks to the pervasive communication and disruptive technologies, it was possible to monitor the vast and rapid spread of the pandemic. Recent experiences with COVID-19 show that a robust and swift approach to collect, process and analyze data and therefore taking precautionary measures helped flatten the pandemic curve. That step eased the burden on the healthcare systems effectively reducing the loss of precious life.

This paper is about leveraging pervasive communication technologies such as the Internet of Things (IoT), Mobile Edge Clouds and, artificial intelligence (AI), as a mechanism for effective management of pandemics [3, 4]. The IoT can be a powerful platform for collecting information such as the location and symptoms of potentially infected individuals [5, 7]. The IoT Cloudlet can provide a platform for various medical devices to connect and relay data from devices to a central service for further processing. These devices include smart thermometers, blood-pressure measurement units, wearable vitals sensors, connected inhalers, and ventilators, etc [19]. Furthermore, smartphone location data can be useful to determine individual's movement patterns etc. This information can be analyzed through sophisticated artificial intelligence algorithms to extract movement patterns of pandemic patients and tracking others who may have come in contact with these patients and may have been potentially infected. This approach is extremely important for healthcare infrastructure to provide treatment to those who need it immediately, and also to limit and slow the spread of the infection [6,7,8].

The IoT and smart devices usually have limited resources on-board and are incapable of processing large amounts of data on their own. Researchers in [28, 29] have developed ways to offload large amounts of data from resource-constrained IoT devices to the cloud for processing. Consequently, the increase in network traffic creates congestion, disrupting the seamless connectivity and affecting the overall quality of service which is ever important in these scenarios. To address these concerns, we develop and deploy an inexpensive and lightweight cluster of Single Board Computers (SBC) that provides cloud services on the Mobile Edge Cloud (MEC). The SBC-MEC cluster provides an opportunity for offloading computation required by applications running sophisticated AI algorithm(s) in the cloud to a nearby SBC-MEC cluster on the edge rather than a distant data center. A prototype of the SBC-MEC cluster with an application for collecting data from medical devices within the IoT is successfully deployed.

The next section of the paper discusses communication technologies that can be utilized to effectively manage pandemics. Section III discusses a platform using pervasive communications technologies and artificial intelligence for managing pandemics. Section IV details the SBC-MEC cluster implementation and deployment. Issues and challenges are presented in section V followed by a summary and conclusions.

## II. PERVASIVE COMMUNICATION TECHNOLOGIES

Over the past few decades, the Internet has evolved rapidly. Smaller and cost-effective hardware components packed with tremendous processing power, have led to many powerful and beneficial applications of the Internet that have changed the way we manage various aspects of our life. These applications are impactful because they provide ease of use, mobile connectivity, broader control, and automation. Hence, an unprecedented number of devices are being connected to the Internet, leading to the evolution of IoT.

IoT is a pervasive form of the Internet that provides connectivity to almost all the objects (things) that we commonly use. These objects are uniquely identifiable, and accessible from anywhere and at any time. These objects are also capable of collecting and/or sensing information, processing it, and communicating it. Technically, an object connected to IoT will have adequate processing power and communication capability. Technological advances continue to make these objects smaller, more powerful, energy-efficient, and cost-effective [10,11]. IoT has rapidly evolved over the past decade and its impact on our daily activities is becoming visible. To put the IoT growth in perspective, the number of devices connected to IoT in 2016 was estimated to be 6.4 Billion, and this number is estimated to be 64 billion by the year 2025. This implies that the average number of devices per capita will be about 8. Such a pervasive level of connectivity can be very effective in collecting data about anyone in the world and even tracing and tracking individuals with the help of IoT devices associated with individuals [12,13]. Such a platform for collecting data and communicating it to a central location such as public cloud platforms [30], is ideal for managing any local, national, or global pandemics.

IoT and artificial intelligence go hand-in-hand for most of the applications. As indicated earlier, IoT is an ideal platform for collecting data. The goal of integrating artificial intelligence with IoT is not only to make devices/systems smart but also to help them make autonomous decisions. Smart devices, equipped with AI capabilities, use available data to learn from past trends and make efficient and effective smart decisions for the future. The data collected by IoT devices can be processed locally (edge computing), or centrally using a cloud computing environment or both. However, due to constrained battery power and the absence of sophisticated AI algorithms onboard IoT devices, the decision-making abilities of edge devices may be limited [14]. Components of an AI environment is depicted in Figure 1 [15].

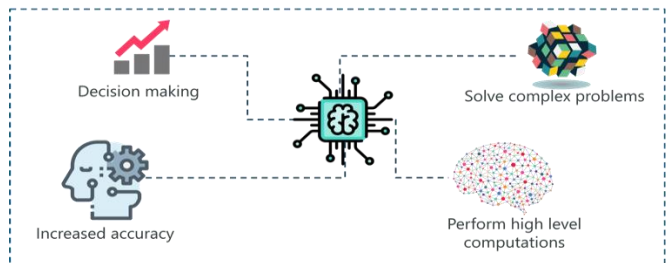

Figure 1. Components of artificial intelligence [15].

## III. MANAGING PANDEMICS BY USING PERVASIVE COMMUNICATION TECHOLOGIES

Before we start discussing the management of pandemics by using pervasive communication technologies, we need to understand how diseases spread to become pandemics and how they can be contained. All pandemics have serious and unfortunate consequences that impact society in many ways resulting in the loss of precious life. The pandemics spread through contact, the speed at which they spread, and the severity of their consequences depend upon several factors including the following [2]:

- Population and population density: Higher the population density, higher will be the rate of spread. There is a higher likelihood of an infected person coming in contact with a larger number of individuals.
- Ways the virus transmits: Some viruses spread through physical contact, some spread through the exchange of bodily fluids; others can be airborne and may spread by an individual being in close proximity of others. The last category is more dangerous than others and is difficult to contain.
- Number of days a virus stays contagious: The duration for which an infected person stays contagious is an important factor. Infected individuals may not even show any symptoms for several days. The only way to find if a person is infected is through testing. A negative test result implies the individual may not be infected that day however, there is no guaranty for that individual to get infected the next day.
- Social dynamics: There is a frequent chance for an individual who visits crowded places or attends a social gathering to become infected than individuals who do not. Precautionary measures such as maintaining good hygiene, washing hands and social distancing are absolutely necessary to reduce the spread of the virus.

World population combined with higher population density in urban regions makes individuals more vulnerable for being infected by others. Social dynamics of a society also play a major role in spreading contagious diseases. In addition, the phenomenon of globalization is a major factor in the rapid spread of pandemics because this phenomenon has shrunk the distances, made travel easier and frequent, has softened the national borders, and has morphed the world into a global village.

The spectacular spread of COVID-19 pandemic has exposed some of the structural deficiencies that otherwise may not have been so obvious. As of June 2020, the number of infected cases continues to grow, the dreadful death toll continues to climb, and the virus continues to spread. These unfortunate facts have left the governments struggling to find ways to effectively contain the spread of COVID-19 and to minimize its impact. A few aspects that are critical for managing this and, for that matter, any pandemic include testing and tracking the carriers and potential carriers of such a virus.

Recognizing a pandemic, identification of infected individuals (by testing), limiting the spread of infection by tracking and implementing appropriate measures/precautions such as social distancing and wearing face masks, are essential to contain and mitigate. Recent experiences with COVID-19 show that a robust and swift approach to stay ahead of the curve can flatten the pandemic curve, avoid overwhelming the healthcare systems, and reduces the loss of precious life.

In managing any pandemic, it is important to understand how it spreads. That will allow developing recommendations for the communities to follow for slowing the spreading rate. For instance, COVID-19 has been spreading by becoming in contact with an infected individual, touching an infected surface, or by being in close proximity to infected individuals. Therefore, the recommendations were developed to wash hands frequently, wear a facial mask, avoid crowded places, and practice social distancing of at least six feet. By practicing these recommendations, the rate of spread can be reduced and that has tremendous advantages. Such interventions flatten the pandemic curve [16]. Flattening the curve gives more time to further understand the nature of the pandemic and find ways to manage it effectively. Having more time will allow the healthcare system and the authorities to prepare and plan for the treatment of patients. In addition, this will allow time to stockpile the necessary quantity of medications to treat pandemic patients and medical equipment including facial masks and personal protective equipment (PPE) for healthcare professionals.

We propose leveraging pervasive communication technologies such as IoT, and AI, as a mechanism for effective management of pandemics [17, 18, 19]. IoT is a very attractive solution for managing pandemics. It can effectively serve as a platform for collecting data about individuals being tested, data about location/movements of infected individuals, and tracing the individuals who may have been infected by being in contact or in close proximity of infected individuals. One of the major advantages of IoT is that any object (things) connected to it, becomes a source of data. Telemetry data originating from these devices can be identified using the MAC address. Each IoT device has its own unique identification and can be accessed from anywhere and at any time. The objects connected to IoT include wearables that continuously monitor vitals (including temperature, oxygen level, heart rate etc.) of the individuals wearing those. In the case of limited testing supplies, this information can become helpful in identifying individuals to be tested. These features provide a powerful platform for collecting information such as location data and symptoms of potentially infected individuals.

This information can be analyzed through sophisticated artificial intelligence algorithms to extract movement patterns of pandemic patients, and tracing others who may have come in contact with these patients and may have been potentially infected. Limited AI algorithms can be implemented by the IoT devices, however the limited resources onboard these devices may not be capable of processing large amounts of data. More advanced analytics can be performed using more powerful computing devices available in remote data-centers through cloud computing [28, 29]. This approach is extremely important for healthcare service provides to extend treatment to individuals in need of immediate attention [20, 21, 22], and also to limit and slow the spread of the infection.

We know that current IoT infrastructure is severely constrained on communication bandwidth. IoT devices are expected to generate copious amounts of data, transferring large amounts of data to the cloud using the existing infrastructure results in network latency and quality of service issues. We address these issues in the next section by deploying a cluster of small, mobile, and low cost on the edge. The integration of a Mobile Edge Cloud (MEC) cluster reduces the network traffic between IoT devices and the cloud by processing and analyzing data locally. The MEC extends the AI application executing in the cloud to the local environment available on the edge of the cloud network.

## IV. SBC-MEC CLUSTERS BASED PROTOTYPE

The SBCs are ever more becoming powerful devices at increasingly lower costs. The low price offered by SBCs is an attractive opportunity for creating SBC based clusters. SBC-MEC clusters provide an opportunity to extend the cloud computing paradigm to the edge at a very low cost, where a mobile micro cloud cluster exists between the data-center and the data source, Figure 2. The data source or IoT devices may consist of sensors and medical devices acquiring data from a location. These devices usually run on batteries having limited onboard resources (processor, memory, etc.) relaying information to a ground station in the field. The ground station would relay the collected data to a centralized cloud-based data center intermittently for further processing.

The SBC-MEC cluster provides an opportunity for offloading computation required by applications running sophisticated AI algorithm(s) in the cloud to a nearby SBC-MEC cluster on the edge rather than a distant data center. Generally, the network latency, quality of service in transmission and computation/service time are important to achieve higher dependability. The SBC-MEC cluster i) processes data accumulated from the IoT devices, and/or ii) processes intermediate data on the edge and later transmit the processed information to the distant data center for further processing or storage. In both cases, the network traffic is reduced, positively affecting the traffic congestion, therefore improving the overall network latency and throughput, which are key characteristics for seamless operability for medical field teams in the event of managing a pandemic.

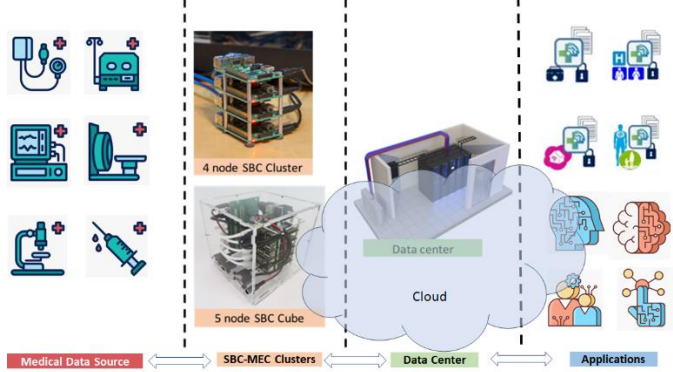

Figure 2. SBC-MEC Clusters providing Cloud services on the edge

### A. Kubernetes as a MEC service orchestrator

The Kubernetes ecosystem deployed on the Edge-cloud leverages the use of lightweight cloud computing technologies suitable for resource constrained SBC devices. The framework utilizes Docker-based lightweight containers [23] with a small footprint in comparison to the traditional virtual machines (VM) that are resource hungry. Most of the operating system functions including file management system, memory address space, and networking can be virtualized in a container. The small size of containers as opposed to a VM allows easier copy and move operations on the container images between devices.

Kubernetes [24] is a container orchestration tool that allows different physical machines to host containers. The distributed environment deploys multiple containers that execute various instances of an application on multiple physical machines. A master node provides a singular point of control providing access control and communication with the slave nodes referred to as Kubernetes agents (KA). A KA is deployed on a physical machine and is responsible for managing the various containers deployed on that physical machine. The Master node is responsible for orchestration consisting of numerous activities including resource allocation, scheduling, container deployment, etc. The Kubernetes orchestrator controls all containers regardless of the physical deployment on the physical nodes.

Figure 3 presents a high-level architecture of the SBC-MEC deployment within the Kubernetes Eco-system. The Master node, execute on the health-care provider/ research center public cloud. The system administrators or users configure and approve the deployment of SBC-MEC cluster requests by medical staff in the field. Once approved, the SBC-MEC clusters configure/deploy containers from the cloud. Various containers can be deployed on physical nodes, within the data center and on the edge in the SBC-MEC. The medical devices/sensors communicate and transfer information from the IoT devices to the cluster using various mediums including but not limited to Application Programming Interfaces (APIs), RESTful web-services, etc.

In the next section, we present the SBC-MEC cluster implementation followed by a RESTful web-service that allows medical devices to connect to the Kubernetes agent, executing the data collection service in a container on the SBC-MEC.

### B. SBC-MEC Clusters

We develop three small clusters using three different kinds of SBCs, Ordoid Xu-4 [25], Raspberry Pi [26] and Latte-Panda [27]. The Ordoid Xu-4 was developed by HardKernel and is available for 49 USD. Xu-4 uses Samsung Exynos5 Quad-core ARM Cortex-A15 Quad 2Ghz and Cortex-A7 Quad 1.3GHz CPUs with 2Gbyte LPDDR3 RAM at 933MHz. Two USB 3.0 ports, as well as a USB 2.0 port, allows faster communication with attached devices.

The Raspberry Pi (RPi) is predominantly the market leader with a low cost and affordable price starting at 23 USD. The RPi Model 3B+ uses a Broadcom BCM2837 System-on-chip (SoC) with a Quad-core processor running at 1.4 GHz. The device allows various modes of connectivity with a Gigabit Ethernet, HDMI, USB, Wi-Fi, and Bluetooth. A newer model 4B is available with 8GB of RAM with a faster 1.5 GHz Quad-core processor.

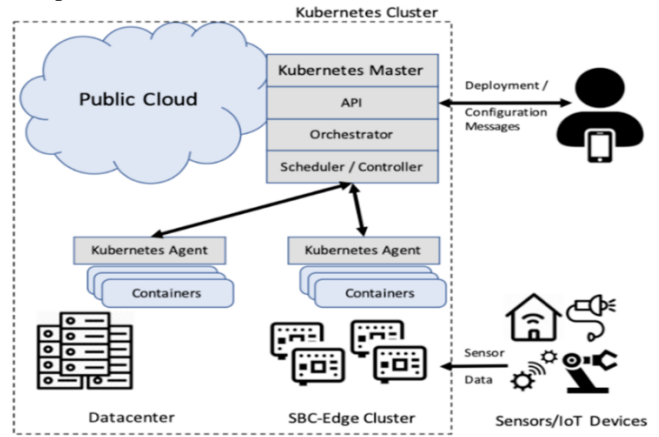

Figure 3. Kubernetes Eco-System for SBC-MEC

The Latte Panda 4G was developed as an SBC device with Microsoft Windows support. This device comes with Intel Atom Cherry Z8350 SoC with a quad core processor at 1.8 GHz, supporting Intel HD Graphics 200-500 GPU and 4 GB of on-board RAM. This device also supports HDMI, Ethernet, Wi-Fi, Bluetooth connectivity, and costs 169 USD. All listed prices are quoted as of June 2020.

SBC based edge clusters using the Raspberry Pi 3B+, Odroid Xu-4, and LattePanda 4G SBCs were built. Figure 4 shows the three clusters with 3 SBC each or Raspberry Pi and Odroid Xu-4 and 2 Latte Pandas in the third cluster. Ubuntu MATE was installed on the clusters built using Raspberry Pi and Xu-4 SBC. Windows 10 IoT was installed on LattePanda. Kubernetes agents/slaves on each device were initialized and configured to connect to the Kubernetes master node. The devices in each cluster connect to a local Ethernet switch that connects to the campus network providing Internet access. To realize the SBC-MEC architecture, we installed the Kubernetes master node on a PC in the lab. This PC connects to the SBC-MEC clusters over the Internet.

## C. Data collection service application

We develop a data collection service application to test the SBC-MEC clusters. The application deployment on the cluster serves as a prototype implementation. The main objective of this application is to log data collected from sensors and medical devices, process the data locally in the SBC-MEC, and send updates to the cloud service.

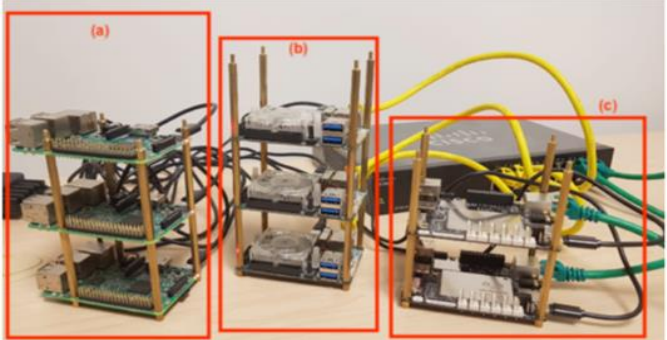

Figure 4. A Kubernetes Agent installed on a 3 clusters. (a) Raspberry Pi (b) Odroid Xu-4 (c) Latte Panda; devices in each cluster connect to a network switch.

The application listens to the `gateway` using `SocketIO` for any updates from within a container. After the initial handshake, the application listens to the `Nodes` event which sends the current status of the connected sensors/medical devices. All received data is appended with a time stamp, the results are stored within the container as log files synchronized by the time stamp. Timestamps are used to calculate the connectivity time of a medical device.

```
socket.on('Nodes', (receivedData) => {
  console.log(data);
  receivedData.data.forEach(MDevice => {
    MDevice.Nodes.forEach(Node => {
      Node['timestamp'] = new Data().getTime();
      Node['connectTime'] = 0;
      NodeList.push(Node);
    });
  });
});

function processData(){
  updateConnectionTime();
  const records = [];
  NodesList.forEach(Node => {
    records.push({NodeID: Node.id, connectionTime:
        .connectionTime});
    Node.connectionTime = 0;
  });
  csvWriter.writeRecords(records)
    .then(() => {
      console.log('Done');
  });
}
```
Figure 5. Listing for `Nodes` script and `processData` function

Another event is `NodesUpdate` which gets triggered at any time the `status` of the connected Node changes. As an update is triggered, the gateway sends relevant information about the selected Node/device. If the status of the Node changes from `connected` to `disconnected`, the difference between the current time-stamp and the stored time stamp is computed, the result is stored in the log file. Similarly, if the status changed from disconnected to connected, the log file is subsequently updated. Every 24 hours, the application reads the log file to process the data. It determines the connectivity time for each device by processing the timestamps of the records annotated as `connected` within the log file. The results are processed into a Comma Separated Value (CSV) format and can be processed further in the cloud by application managers. The `processData` function is called at regular intervals which can be specified by the user. Figure 5 shows the `Nodes` script and the `processData` function used in the application.

The SBC-MEC cluster was deployed with-in the lab environment. The results demonstrate the accurate functioning of the prototype.

## V. ISSUES AND CHALLENGES

Integrating IoT, Cloud and AI solutions to create an effective and efficient platform for managing pandemics is expected to result in the following:

1. A robust and smart platform for managing pandemics such as COVID-19. The platform will have capabilities for identifying individuals for testing and tracking their movement. IoT can help with tracking the movement of infected individuals.
2. A smart platform using AI for forecasting the growth patterns of the virus after integrating the identified location of infected or potentially infected individuals with geographical maps and population maps.
3. A platform that is scalable and applicable to city, county, state, or country.

There are various issues and challenges to developing such a smart platform.

The world population is expected to be about 10 Billion by 2050 and naturally, population clusters will have higher population density. In pervasive communication environments, it is expected that there will be an average of eight connected (and mostly mobile) devices per capita. Each of these devices will have its own, globally unique internet address to receive and send information. The effectiveness and efficiency of such a system for managing pandemics is highly dependent on the available resources. We know that current IoT infrastructure is severely constrained on communication bandwidth. IoT devices have limited processing capability and limited energy resources.

The population density will impact the analytical outcomes and will influence the measures to be employed for containment and mitigation of a pandemic. In addition to IoT telemetry data originating from medical devices and wearables, testing and tracking data of individuals with the geographical mapping along with the population data will be considered. Deployment of these solutions in the densely populated areas will generate copious amounts of data that may overwhelm the infrastructure.

Striking a balance between the amount of AI data analysis at the IoT edge devices and clusters, and the amount of data offloaded to the cloud for further analysis is also an important aspect to consider. Such a balance depends on many factors including the processing power available at the IoT edge devices, the energy (and hence the operational longevity) of the edge devices, the complexity of the decisions, and nature of the decisions (local or broader), etc. The efficiency of the AI data analysis is not expected to have a linear relationship with the division of data processing between the edge devices and the cloud. For a given situation, there must be an optimal division of data processing tasks among the edge devices and the cloud.

Maintaining the privacy of information is essential and it is assumed that the system will have provisions for safeguarding the personal information of individuals.

Overall, for a smart and robust approach to managing pandemics, robust communication, and computing infrastructure is absolutely essential.

## VI. Summary and Conclusions

Pandemics are real and historically have had serious consequences unless they were effectively contained. Management of pandemics requires the identification of infected individuals (by testing), limiting the spread of infection by tracking and implementing appropriate measures/precautions such as social distancing and using personal protective equipment. Recent experiences with COVID-19 show that by using a smart and swift approach to deal with pandemics, avoids overwhelming of healthcare systems, and reduces the loss of precious life. This paper has proposed the use of pervasive communication technologies such as IoT, MEC and AI, as an approach to effectively manage pandemics. IoT provides pervasive connectivity among the common devices and can be used for collecting information such as the location and symptoms of potentially infected individuals. This information can be offloaded to a MEC cluster or the cloud where it can be analyzed through sophisticated AI algorithms to extract useful information. An inexpensive and lightweight cluster of SBC is deployed that provides cloud services on the MEC. A prototype of the SBC-MEC cluster with an application for collecting data from medical devices within the IoT is successfully deployed. This approach is essential for managing pandemics by avoiding the possibility of overwhelming the healthcare system. The platform can also be effective in containing the spread of the infection.